\documentclass[journal]{IEEEtran}
\usepackage{algorithm}
\usepackage{moreverb}
\usepackage{amsmath}
\usepackage{colortbl}
\usepackage{wrapfig}
\usepackage{bm}
\usepackage{algorithmic}
\usepackage{tabularx}
\usepackage{cite}
\usepackage{graphicx}
\usepackage{amsfonts}
\usepackage{amsthm}
\usepackage{amssymb}
\usepackage{cases}
\usepackage{bbm}
\usepackage{booktabs}
\usepackage{float}
\usepackage{stfloats}
\usepackage{subfigure}
\usepackage{xcolor}
\usepackage{colortbl}
\usepackage{xpatch}
\usepackage{epstopdf}
\usepackage{algorithm}
\usepackage{algorithmic}
\usepackage{setspace}

\makeatletter
\ExplSyntaxOn
\cs_new:Npn \bibColoredItems #1#2
{
	\clist_map_inline:nn {#2} { \cs_new:cpn {bib@colored@##1} {#1} } 
}
\ExplSyntaxOff

\newcommand\bib@setcolor[1]{%
	\ifcsname bib@colored@#1\endcsname
	\expandafter\color\expandafter{\csname bib@colored@#1\endcsname}
	\else
	\normalcolor
	\fi
}

\xpatchcmd\@bibitem
{\item}
{\bib@setcolor{#1}\item}
{}{\fail}

\xpatchcmd\@lbibitem
{\item}
{\bib@setcolor{#2}\item}
{}{\fail}
\makeatother

\ifCLASSOPTIONcompsoc
\usepackage[caption=false,font=normalsize,labelfont=sf,textfont=sf]{subfig}
\else
\usepackage[caption=false,font=footnotesize]{subfig}
\fi
\ifCLASSOPTIONcaptionsoff
\usepackage[nomarkers]{endfloat}
\let\MYoriglatexcaption\caption
\renewcommand{\caption}[2][\relax]{\MYoriglatexcaption[#2]{#2}}
\fi
\usepackage{color}

\allowdisplaybreaks[4]
\begin{document}
		
	\title{\huge 
Scalable Design for RIS-Assisted Multi-User {Downlink} System Empowered by RSMA under Partial CSI
}
	\allowdisplaybreaks
	\author{ Yifan Fang,~Bile Peng,~\IEEEmembership{Senior Member,~IEEE}, Yingyang Chen,~\IEEEmembership{Senior Member,~IEEE},\\Qiang Li,~\IEEEmembership{Senior Member,~IEEE},~Marwa Chafii,~\IEEEmembership{Senior Member,~IEEE}, Eduard A. Jorswieck,~\IEEEmembership{Fellow,~IEEE}\\
    \thanks{Y. Fang, Y. Chen, and Q. Li are with the Department of Electronic Engineering, College of Information Science and Technology, Jinan University, Guangzhou 510632, China (e-mail: loey2493@stu2023.jnu.edu.cn; chenyy@jnu.edu.cn; qiangli@jnu.edu.cn).}
    \thanks{B. Peng and E. A. Jorswieck are with the Institute for Communications Technology, TU Braunschweig, Germany (e-mail: b.peng@tu-braunschweig.de; e.jorswieck@tu-braunschweig.de).}
    \thanks{M. Chafii is with the Wireless Research Lab, Engineering Division, New York University Abu Dhabi, UAE. She is also with NYU WIRELESS, NYU Tandon School of Engineering, New York, USA (e-mail: marwa.chafii@nyu.edu)}}
		\vspace{-5mm}

	\maketitle

\begin{abstract}
In large-scale reconfigurable intelligent surface (RIS) communication systems, the precise acquisition of channel state information (CSI) is challenging. {Consider} a practical RIS configuration where only a few reflective elements serve as anchors to estimate CSI, which are termed partial CSI. To improve the robustness against partial CSI and the scalability of RIS networks, this paper proposes an unsupervised learning-based rate-splitting multiple access (RSMA) scheme for RIS-assisted multi-user systems. Specifically, {RISnet, a neural network architecture designed to infer full CSI from partial observations, is employed and integrated} with a low-complexity RSMA precoder. Effective channel features are constituted from partial CSI, and the original elements with channel information contribute to new anchors after expansion in RISnet. Numerical results demonstrate that the proposed scheme approximates the performance with a full CSI of RIS under deterministic {ray-tracing} channel conditions. When channel uncertainty increases during training, RSMA has been shown to enhance RISnet robustness, significantly mitigating performance loss. 

\end{abstract}
\vspace{-1mm}
\begin{IEEEkeywords}
Rate-splitting multiple access (RSMA), reconfigurable intelligent surface (RIS), channel uncertainty, RISnet, unsupervised learning.
\end{IEEEkeywords}
\vspace{-3mm}
 
	\vspace{-1mm}
    
	\section{Introduction}
	\vspace{-1mm}

\begin{spacing}{0.99}
Reconfigurable intelligent surface (RIS) has emerged as a key technology for next-generation mobile communications~\cite{zhuo2023ris}. It can intelligently {re-engineer the wireless propagation environment}, converting
non-line-of-sight (NLoS) links to virtual line-of-sight (LoS) ones. Reaping the benefits of RIS heavily relies on the acquisition of channel state information (CSI) \cite{peng2024deep}. However, precisely acquiring full CSI is challenging, especially for large-scale RISs given the large number of reflective units~\cite{wang2020channel}.

Recently, rate splitting has emerged as a promising strategy for future wireless networks~\cite{clerckx2016rate}. Using precoding and rate splitting at the transmitter, together with successive interference cancellation (SIC) at receivers, rate splitting multiple access (RSMA) offers flexible interference management~\cite{mao2018rate}. In particular, it has been shown to offer greater generality and robustness than space-division multiple access (SDMA) and non-orthogonal multiple access (NOMA).

Recent studies have explored the integration of RIS and RSMA, showing the potential gains in the aspect of resource allocation~\cite{fu2021resource}, outage probability~\cite{bansal2022rate}, and spectral efficiency~\cite{zhao2022ris}.  To simplify the analysis, most studies assume perfect CSI\cite{fu2021resource,bansal2022rate,zhao2022ris}. With the advancement of neural {network–based methods}, researchers have utilized machine learning~\cite{nguyen2021machine}, deep learning \cite{jiang2021learning}, and deep reinforcement learning~\cite{feng2020deep} to predict channels and optimize RIS configurations. {However, the complexity surges as the size of RIS scales up.} 

\vspace{-0.5mm}The scalability of RIS is another key challenge, referring to the ability of the system to maintain desired performance as the size of RIS enlarges. 
Peng \emph{et al.}\cite{peng2023risnet} proposed a neural network architecture named RISnet, 
whose parameter count is independent of the RIS scale. However, its effectiveness depends on the specific channel model and requires accurate partial CSI at the base station (BS). Notably, {RSMA was} shown to exhibit robustness to imperfect CSI due to its flexible interference management\cite{wang2025autonomous,zhu2024rate}. This inspires us to integrate RISnet to support a scalable RIS network intrinsically, particularly under complex channel conditions, despite the paucity of results at the time of writing. 

We consider a practical RIS where only a few reflective units serve as anchors to estimate CSI via pilot signals, while the majority remain fully passive, resulting in a simplified hybrid RIS without signal amplification \cite{ju2024beamforming}. {The channel gain between RIS and the user acquired at the anchors is termed partial CSI}. When the channel contains strong propagation paths (e.g., LoS or specular ones), the obtained partial CSI can be sufficient to infer the full CSI~\cite{peng2023risnet}.
Building on this, we propose a scalable communication scheme based on the RIS-assisted RSMA philosophy, even with partial CSI. The contributions are threefold:
\begin{itemize} 
\item[1)] {We {consider} a RIS-assisted multi-user RSMA downlink system {with partial CSI acquisition.}} Subject to unit modulus and power constraints, we formulate a weighted sum rate (WSR) maximization problem to optimize the precoding matrix and phase shift configurations.

	\item[2)] We propose a hybrid solution that integrates the proposed low-complexity RSMA precoder and RISnet training. We disclose how to obtain an effective channel feature from partial CSI and the overall architecture of RISnet. The scheme has high scalability due to the input of RISnet being independent of the RIS size.
 
    	\item[3)]Simulation results demonstrate that the performance of the proposed scheme approximates to a full CSI counterpart under deterministic conditions. When partial CSI is insufficient to derive full CSI under fully random channel conditions, RSMA enhances robustness, significantly mitigating performance loss.
\end{itemize}
\end{spacing}
\vspace{-1mm}
\section{System Model and Problem Formulation }
\setlength{\abovecaptionskip}{-0.05cm}
\setlength{\belowcaptionskip}{-0.2cm}
\label{Sec_SysMod}
\begin{spacing}{0.95}
We consider a scenario where a BS transmits downlink RSMA signals to multiple users with the aid of RIS. 
This system comprises the BS equipped with $M$ antennas, $K$ single-antenna users, and one large-scale RIS with $N$ reflective units. ($N\gg M$). For the RIS, we assume only $N'$ reflective units are fitted with RF chains($N'\ll N$). The user set is denoted by $\mathcal{K} = \left\{ {1,...,K} \right\}$, where $K\leqslant M$. With the 1-layer RSMA scheme at BS, the message $ W_k $ sent to user $k$ is split into the private part $ W_{p,k} $ and the common part $ W_{c,k} $.  Then, all common parts $\{ {W_{c,1}},...,{W_{c,K}}\} $ are encoded into a single common stream ${s_c}\in \mathbb{C}$, while the private part of the $ {k}^{th}$ user is independently encoded into the private stream ${s_k}\in \mathbb{C}$. 
BS transmits the signal ${\bf{s}}=[s_c,s_1,...,s_K]^T$ using the precoding matrix ${\bf{V}} = [{\mathbf{v}}_c,{\mathbf{v}}_1,...,{\mathbf{v}}_K]$, where ${\mathbf{v}}_c$, ${{\mathbf{v}}_k}\in {\mathbb{C}^{M \times 1}}$ are precoding vectors applied to the common and the $k$-th private stream, respectively. 
The transmit signal is

\begin{small}
\begin{align}
	\label{transmit signal}
	{\mathbf{x}} = {\bf{Vs}}= {{\mathbf{v}}_c}{s_c} + \sum\nolimits_{k \in {\mathcal K}}  {\mathbf{v}}_k{s_k}, 
\end{align}\end{small}where $\mathbb{E}\{\mathbf{s}{\mathbf{s}^H}\} = {\mathbf{I}}$, and the transmit power budget is ${P_t}$, i.e., ${\rm{tr}}({\bf{V}}{{\bf{V}}^H})\le{P_t}$. The received signal at the users is
\begin{small}
\begin{align}
	\label{received signal}
{\bf{y = (G\Phi L + D)x + n = Hx + n}},
\end{align}\end{small}where ${\bf{D}}\in {\mathbb{C}^{K\times M}}$, ${\bf{L}}\in {\mathbb{C}^{N \times M}}$, and ${\bf{G}}\in {\mathbb{C}^{K\times N}}$ denote the transmission channels from BS to users, from BS to RIS, and from RIS to users, respectively. The diagonal matrix ${\bf{\Phi}}\in {\mathbb{C}^{N \times N}}$ is the phase shift configuration matrix of RIS, with $\varphi_{nn}$ $(n\in\{1,\cdots,N\})$ denoting the $n$-th diagonal element. ${\bf{n}}=[n_1,...,n_K]^T\in {\mathbb{C}^{K\times 1}}$ is the thermal noise vector, where $n_k$ is the part perceived at user $k$ with the power of $\sigma _k^2$. {Besides, ${\bf{H=G}\bf{\Phi}\bf{L}+\bf{D}}=[{\bf{h}}_1,...,{\bf{h}}_K]^T\in {\mathbb{C}^{K\times M}}$ is the combined transmission channel matrix, where  ${\bf{h}}_k\in {\mathbb{C}^{M\times 1}}$.} 
At the receiver side, each user first decodes the common stream $s_c$ and removes it using the SIC technique. Then, user $k$ decodes its own private stream $s_k$, while other private ones are treated as interference. Specifically, the signal-to-interference-plus-noise ratio (SINR) of the common stream and the private stream at user $k$ are respectively given by

\begin{small}
\begin{align}
	\label{sinr_c}
{{\rm{SINR}}_{c,k}} = \frac{{{\left| {{\bf{h}}_k^T{{\mathbf{v}}_c}} \right|}^2}}{\sum\nolimits_{i \in {\mathcal K}}{{{\left| {{\bf{h}}_k^T{{\mathbf{v}}_i}} \right|}^2} + \sigma _k^2} },
\end{align}\end{small}
	\vspace{-2mm}
\begin{small}\begin{align}
	\label{sinr_p}
{{\rm{SINR}}_{p,k}} = \frac{{\left| {{\bf{h}}_k^T{{\mathbf{v}}_k}} \right|}^2}{\sum\nolimits_{i\in {\mathcal K},i\ne{k}}{{{\left| {{\bf{h}}_k^T{{\mathbf{v}}_i}} \right|}^2} + \sigma _k^2} }.
\end{align}\end{small}The instantaneous rate at user $k$ for decoding ${s_c}$ and ${s_k}$ are $R_{c,k}= {\log _2}(1 + {{\rm{SINR}}_{c,k}})$ and $R_k={\log _2}(1 + {{\rm{SINR}}_{p,k}})$, respectively. To guarantee the common stream ${s_c}$ can be decoded by all users, the achievable common rate is ${R_c}=\min_{k\in\mathcal{K}} \{ {R_{c,k}}\}$. Since ${R_c}$ is shared by all users, we have $\sum\nolimits_{k\in\mathcal{K}} {{C_k}}={R_c}$, where ${C_k}$ is the common rate portion of user $k$. For simplicity, we assume that $R_c$ is evenly allocated to each user, i.e., $C_k=R_c/K$\cite{dizdar2021rate}. Thus, the achievable rate of user $k$ is $R_{k,tot}=\min_{k\in\mathcal{K}} \{{R_{c,k}}\}/K+R_k$. 
The objective is to improve the overall transmit rate while satisfying the power budget constraints, the physical characteristics of the RIS reflective units, and the decoding requirements for the common stream. Explicitly, we consider jointly optimizing the precoding matrix $\bf{V}$ and the phase shift matrix $\bf{\Phi}$ since their optimal values depend on each other, and construct the WSR maximization problem as

\begin{small}
\begin{subequations}
	\label{P1}
	\begin{align}
		&\begin{gathered}
			\hspace{-8mm}\mathcal{P}_1: \mathop {\max }\limits_{{\bf{V}},\bf{\Phi}}\quad	\sum\nolimits_{k\in\mathcal{K}}   {{a_k}R_{k,tot} } \hfill \label{P1a}
		\end{gathered} \\
		&\hspace{0mm}{s.t.\hspace{3mm}{\rm{tr}}({\bf{V}}{{\bf{V}}^H})\le{P_t}, }\label{P1b}\\
      &\hspace{8mm}{\left|{\varphi_{nn}}\right|=1,n\in \left\{ {1,...,N} \right\}.}\label{P1c}
	\end{align}
\end{subequations}\end{small}Here, ${a_{k}>0}$ denotes the rate weight of user $k$, (5b) specifies the transmit power budget, and (5c) is a unit-modular constraint related to the RIS. The problem \eqref{P1} is difficult to solve. Thus, we propose an unsupervised learning (UL) approach based on RISnet that integrates the weighted minimum mean square error (WMMSE) precoding scheme to reduce the solution complexity.
\end{spacing}

\vspace{-3mm}
\section{The Proposed Approach with RISnet and RSMA}

\begin{spacing}{0.95}
In this section, we describe how to solve the formulated problem via RISnet training and the proposed RSMA precoder. Then, we disclose how to obtain an effective channel feature from partial CSI and the overall architecture of RISnet. 
\vspace{-2mm}
\subsection{RISnet Training}
We propose a two-stage UL approach to solve the problem. In the offline training stage, a neural network ${{N}_{\theta}}$ is trained to learn the mapping between available partial CSI, the phase shift matrix $\bf{\Phi}$, and the precoding matrix $\bf{V}$. To reduce the training difficulty and exploit the high performance of existing analytical precoding schemes, we update $\bf{V}$ and $\bf{\Phi}$ alternatively in this stage. Since the joint otimization of precoding and RIS configuration is an open problem, it would be infeasible to acquire the correct labels if we decide for supervised learning. Therefore, we choose unsupervised learning to let the neural network look for the optimal solution by itself. More specifically, we have ${\bf{\Phi}} = {{N}_{\theta}}({\bf{\Gamma}})$, where ${\theta}$ is the trainable parameter of ${{N}_{\theta}}$, and ${\bf{\Gamma}}$ is the channel feature constituted from the partial CSI of RIS, which will be explained in Section III-C. Then, in the online deployment stage, the trained ${{N}_{\theta}}$ directly maps the observed partial CSI to the optimized beamformer and RIS configuration.

Both the training and test datasets are generated by simulating random samples from the system model described in Section II. After initializing RISnet, a batch of data samples is randomly selected. For this batch of data samples, the precoding matrix $\bf{V}$ corresponding to the current RIS configuration is computed and fixed for the subsequent training process. Section III-B elaborates on the obtaining of the proposed precoder. 
The objective function of \eqref{P1} is then expressed as ${{f}({\bf{\Gamma}},{\bf{\Phi}})={f}({\bf{\Gamma}},{N}_{\theta}({\bf{\Gamma}});{\theta})}$, depending on ${\bf{\Gamma}}$ and $\bf{\Phi}$. Thus, the learning problem can be described as

\begin{small}\begin{equation}\label{learning problem}
{\max \limits_{\theta}  \sum\nolimits_{{\bf{\Gamma}} \in {\mathcal D}} {{f}({\bf{\Gamma}},{N}_{\theta}({\bf{\Gamma}});{\theta})}}, 
\end{equation}\end{small}where $\mathcal D$ represents the training dataset. Next, compute the phase shift matrix $\bf{\Phi}$ with current ${{N}_{\theta}}$ and then calculate the objective function. Subsequently, by calculating the gradient of the objective function with respect to the neural network, a stochastic gradient ascent step is performed to optimize ${\theta}$. This process is repeated with a new batch of data samples until convergence or achieving the predefined iteration number. In this way, we train RISnet without providing labels. Algorithm 1 details the training procedures of RISnet.
\vspace{-3mm}

\subsection{RSMA Precoder}
In this part, we design a WMMSE precoder for RSMA as a low-complexity analytical scheme, which is utilized to ensure precoding performance and reduce training overhead without extra precoder optimization. As shown in Algorithm 1, when calculating the precoding matrix $\bf{V}$ on the BS side, we treat the current phase shift matrix $\hat{\bf{\Phi}}$ as constant. Consequently, we obtain the estimated combining matrix $\hat{\bf{H}}=[\hat{\bf{h}}_1,\cdots,\hat{\bf{h}}_K]^T\in {\mathbb{C}^{K\times M}}$.  Then, problem $\mathcal{P}_1$ is reformulated as

\begin{small}\begin{subequations}
	\label{P2}
	\begin{align}
		&\begin{gathered}
			\hspace{-8mm}\mathcal{P}_2: \mathop {\max }\limits_{\bf{V}}\quad	\sum\nolimits_{k\in\mathcal{K}}   {{a_k}(\min_{k\in\mathcal{K}}\{{R_{c,k}}\}/K+R_k) } \hfill \label{P2a}
		\end{gathered} \\
		&\hspace{0mm}{s.t.\hspace{3mm}{\rm{tr}}({\bf{V}}{{\bf{V}}^H})\le{P_t}. }\label{P2b}
	\end{align}
\end{subequations}\end{small}

At user $k$, let $\hat{s}_{c,k}=g_{c,k}y_k$ and $\hat{s}_k=g_{k}(y_k-\hat{\bf{h}}_k{{\mathbf{v}}_c}s_c)$ be the estimates of common stream $s_{c,k}$ and private stream $s_k$, respectively, where $g_{c,k}$ and $g_k$ denote the corresponding equalizers. The MSEs for decoding the common and private streams are formulated by $\varepsilon_{c,k}=\mathbb{E}\{{\left|s_c-\hat{s}_{c,k}\right|^2}\}$ and $\varepsilon_k=\mathbb{E}\{{\left|s_k-\hat{s}_k\right|^2}\}$, and can be calculated as

\begin{small}\begin{align}
	\label{MSEs}
	&\varepsilon_{c,k}={\left|g_{c,k}\right|^2}T_{c,k}-2\Re(g_{c,k}\hat{\bf{h}}_k{\mathbf{v}}_c)+1, \notag\\
    &\varepsilon_k={\left|g_k\right|^2}T_k-2\Re(g_k\hat{\bf{h}}_k{\mathbf{v}}_k)+1,
\end{align}\end{small}
where $T_{c,k}=\sigma _k^2+\sum\nolimits_{i\in\mathcal{K}}\left|\hat{\bf{h}}_k{\mathbf{v}}_i\right|^2+{\left| {\hat{\bf{h}}_k{{\mathbf{v}}_c}} \right|}^2$ and $T_k=T_{c,k}-{\left| {\hat{{\bf{h}}_k}{{\mathbf{v}}_c}} \right|}^2$ are the received power and the power after removing the common stream at user $k$, respectively. By solving $\frac{\partial \varepsilon_{c,k}}{\partial g_{c,k}}=0$ and $\frac{\partial \varepsilon_k}{\partial g_k}=0$, the optimum minimum MSE equalizers are
\vspace{-1mm}
\begin{small}\begin{align}
	\label{MMSE}
	g_{c,k}^{\rm{MMSE}}={\mathbf{v}}_c^H\hat{\bf{h}}_k^H(T_{c,k})^{-1},\space  g_k^{\rm{MMSE}}={\mathbf{v}}_k^H\hat{\bf{h}}_k^H(T_k)^{-1}.
\end{align}\end{small}Thus, minimum MSEs are denoted by
\vspace{-1mm}
\begin{small}\begin{align}
	\label{MMSEs}
	\varepsilon_{c,k}^{\rm{MMSE}}=T_{c,k}^{-1}I_{c,k},\space
    \varepsilon_k^{\rm{MMSE}}=T_k^{-1}I_k,
\end{align}\end{small}where $I_{c,k}=T_k$ and $I_k=I_{c,k}-{| {\hat{\bf{h}}_k{{\mathbf{v}}_k}} |}^2$. Then, the achievable rates can be given by $\hat{R}_{c,k}=-{\log_2}(\varepsilon_{c,k}^{\rm{MMSE}})$ and $\hat{R}_k=-{\log_2}(\varepsilon_k^{\rm{MMSE}})$. The augmented weighted MSEs are
\begin{small}\begin{align}
	\label{Weighs}
	\xi_{c,k}=u_{c,k}\varepsilon_{c,k}-{\log_2}(u_{c,k}),\space  \xi_k=u_k\varepsilon_k-{\log_2}(u_k),
\end{align}\end{small}where $u_{c,k}$ and $u_k$ are the weights associated with each stream at user $k$. By solving $\frac{\partial \xi_{c,k}}{\partial g_{c,k}}=0$ and $\frac{\partial \xi_k}{\partial g_k}=0$, we can express the optimum weight as
\begin{small}
\begin{align}
	\label{Weights}
	u_{c,k}^{\rm{MMSE}}=(\varepsilon_{c,k}^{\rm{MMSE}})^{-1},\space  u_k^{\rm{MMSE}}=(\varepsilon_k^{\rm{MMSE}})^{-1}.
\end{align}\end{small}

Substituting (10) and (12) into (11), we can establish the rate-WMSE relationship as $\xi_{c,k}=1-\hat{R}_{c,k}$ and $\xi_k=1-\hat{R}_k$. Thus, problem \eqref{P2} can be formulated as

\begin{small}
\begin{subequations}
\vspace{-1.5mm}
	\label{P3}
	\begin{align}
		&\begin{gathered}
			\hspace{-8mm}\mathcal{P}_3: \mathop {\min }\limits_{\bf{V,u,g}}\quad	\sum\nolimits_{k\in\mathcal{K}}   {{a_k}(\max\{{\xi_{c,k}}\}/K+\xi_k) } \hfill \label{P3a}
		\end{gathered} \\
		&\hspace{0mm}{s.t.\hspace{3mm}{\rm{tr}}({\bf{V}}{{\bf{V}}^H})\le{P_t}, }\label{P3b}
	\end{align}
\end{subequations}\end{small}where ${\bf{u}}=[u_{c,1},\cdots,u_{c,K},u_1,\cdots,u_K]$ is the weight vector and ${\bf{g}}=[g_{c,1},\cdots,g_{c,K},g_1,\cdots,g_K]$ is the equalizer vector. To solve problem $\mathcal{P}_3$, a common approach is to alternatively optimize $\bf{V}$, $\bf{u}$, and $\bf{g}$ via CVX toolbox until convergence. However, the iterative calculations in CVX increase training time. To further reduce computational overhead, we relax $\mathcal{P}_3$ by replacing the objective with ${\sum\nolimits_{k\in\mathcal{K}}  {{a_k}(\xi_{c,k}}+\xi_k)}$ and using the iterative WMMSE approach introduced in \cite{shi5946304}, which allows us to update $\bf{V}$ by

\begin{small}
\begin{equation}
\left\{
\begin{aligned}
{\mathbf{v}}_c & =  ({\bf{A}}+\mu{\bf{I}})^{-1}\sum\nolimits_{k\in\mathcal{K}}a_k\hat{\bf{h}}_k^Hu_{c,k}g_{c,k},\\
{\mathbf{v}}_k & = ({\bf{A}}+\mu{\bf{I}})^{-1}a_k\hat{\bf{h}}_k^Hu_kg_k.\\
\end{aligned}
\right.
\end{equation}\end{small}where $\mu\ge 0$ is selected via {bisection method} to meet power budget, ${\bf{A}}=\sum\nolimits_{k\in\mathcal{K}}(a_k\hat{\bf{h}}_k^H{\bf{u}}_{c,k}{\bf{g}}_{c,k}{{\bf{u}}_{c,k}}^H\hat{\bf{h}}_k+a_k\hat{\bf{h}}_k^H{\bf{u}}_k{\bf{g}}_k{{\bf{u}}_k}^H\hat{\bf{h}}_k)$. Thus, for problem $\mathcal{P}_3$, $\bf{g}$, $\bf{u}$, and $\bf{V}$ can be alternatively updated by calculating formulas (9), (12), and (14) directly, much decreasing the complexity. To satisfy power constraints in initialization, we set ${\mathbf{v}}_c={P_c}\hat{\mathbf{v}}_c$ and ${\mathbf{v}}_k={P_p}\hat{\mathbf{v}}_k$ with ${P_c}=(1-t){P_t}$ and ${P_p}=t{P_t}/K$. 
$t\in(0,1]$, and $\hat{\mathbf{v}}_c$, $\hat{\mathbf{v}}_k$ are normalized vectors constructed from $\hat{\bf{H}}$.

\begin{algorithm}[t]
\caption{RISnet Training}\label{alg:1}
\begin{algorithmic}[1]
\setstretch{0.9} 
\STATE Randomly initialize RISnet $N_\theta$.
\STATE \textbf{repeat}
\STATE \quad Randomly select a batch of data samples.
\STATE \quad Compute WMMSE precoding matrix ${\bf{V}}$ with current RIS configuration.
\STATE \quad Compute ${\bf{\Phi}} = N_\theta({\bf{\Gamma}})$ for every data sample.
\STATE \quad Compute the combined matrix $\textbf{\bf{H}}$ for every data sample.
\STATE \quad Compute the objective function (5a) for every sample.
\STATE \quad Compute the gradient of (5a) with respect to $\theta$.
\STATE \quad Update the model {with} a stochastic gradient ascent step.
\STATE \textbf{until} Convergence or a predefined iteration limit is met.
\end{algorithmic}
\end{algorithm}
\vspace{-2mm}
 
	\vspace{-2mm}
	\subsection{Construct Channel Feature from Partial CSI}
	\vspace{-1mm}
In this part, we elaborate on how to obtain a channel feature ${\bf{\Gamma}}$ from partial CSI as an effective input to RISnet. Specifically, the channel from BS to RIS $\bf{L}$ is assumed to be obtained since the LoS dominates and the positions of BS and RIS are fixed, {while the NLoS components are sufficiently weak to be negligible}. In contrast,  the BS to users channel matrix $\bf{D}$ and the RIS to users channel matrix $\bf{G}$ vary with the user position. 
Since only $N'$ reflective units fitted with RF chains can estimate CSI, $\bf{G}$ is sparse at first. More specifically, define ${\bf{J}}={\bf{DL^{\dagger}}}$, where $\bf{L^{\dagger}}$ is the pseudo-inverse of $\bf{L}$. We rewrite \eqref{received signal} as ${\bf{y = (G\Phi + J)Lx + n}}$. Explicitly, we define the channel feature of user $k$ coherent with reflective unit $n$ as $\gamma_{kn}=[\left|g_{kn}\right|,{\rm{arg}}(g_{kn}),\left|j_{kn}\right|,{\rm{arg}}(j_{kn})]^T\in {\mathbb{R}^{4\times 1}}$ {to encapsulate all necessary channel information}. Here, $g_{kn}$ and $j_{kn}$ represent the element on $k$-th row and $n$-th column of $\bf{G}$ and $\bf{J}$, respectively. The notation $\left|\cdot\right|$ and $\rm{arg}(\cdot)$ denote the amplitude and phase of a complex number, respectively. The complete channel feature ${\bf{\Gamma}}\in {\mathbb{R}^{4\times K\times N}}$ is a three-dimensional tensor, with the element indexed by $k$ and $n$ in the $2{nd}$ and $3{rd}$ dimensions being $\gamma_{kn}$.
 
\vspace{-3mm}\subsection{Network Architecture}
\begin{figure}[t]
	\centering
	\scalebox{0.35}
	{\includegraphics[width=7.2in]{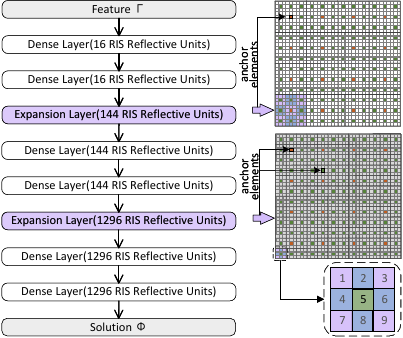}}
	\caption{The RISnet architecture with channel feature $\Gamma$ as input and trained RIS configuration $\Phi$ as output, where the information processing of dense layers is given by (15) while that of expansion layers is given by (16).}
	\label{Fig_flowchart}
	\vspace{-5mm}
\end{figure}

The RISnet architecture is designed for a RIS with a few units capable of estimating channels from pilot signals from users.
This concept is comparable to hybrid RIS~\cite{ju2024beamforming},
where a few active units can amplify signals,
but it is simpler
because the units only estimate the channel, not amplify signals.
The RISnet architecture is shown in Fig.~\ref{Fig_flowchart}, with channel feature as input and consists 8 layers including 6 dense and 2 expansion ones. In fact, we can tailor the number and location of expansion layers, as well as the total number of layers $L$ within the network, to meet the specific requirements.

Since the optimal phase shift matrix of a reflective unit depends on the local information of the current reflective unit and the current user, as well as the global information of all reflective units and different users, we categorized all users and available reflective units into four classes:
\begin{itemize}
\item $cc$: current user and current RIS reflective unit,
	
\item $ca$: current user and all usable RIS reflective units,

\item $oc$: other users and current RIS reflective unit,

\item $oa$: other users and all usable RIS reflective units.
\end{itemize}

Before dealing with the dense layer, we define the input feature of user $k$ and RIS reflective unit $n$ in layer $i$ as ${\bf{f}}_{kn,i}$, where $i<L$ (in particular, ${\bf{f}}_{kn,1}=\gamma_{kn}$),  and use the output stack as input for the subsequent layer. If layer $i$ is a dense layer, the output feature of user $k$ and reflective unit $n$ of layer $i$ is calculated as
	\vspace{-1mm}
\begin{small}
\begin{align}
	\label{dense layer's feature}
    \begin{gathered}
{\bf{f}}_{kn,i+1}=\hfill\\\left(\begin{array}{c}\text{ReLU}(\mathbf{W}_i^{cc} \mathbf{f}_{kn,i} + \mathbf{b}_i^{cc})  \\
\frac{1}{N} \sum_{n'} \text{ReLU}(\mathbf{W}_i^{ca} \mathbf{f}_{kn',i} + \mathbf{b}_i^{ca}) \\
\frac{1}{K-1} \sum_{k' \neq k} \text{ReLU}(\mathbf{W}_i^{oc} \mathbf{f}_{k'n,i} + \mathbf{b}_i^{oc}) \\
\frac{1}{N(K-1)} \sum_{k' \neq k} \sum_{n'} \text{ReLU}(\mathbf{W}_i^{oa} \mathbf{f}_{k'n',i} + \mathbf{b}_i^{oa})\end{array}\right),
	\end{gathered}
\end{align}
\end{small}where  ${{\bf{b}}^{cc}_i}\in {\mathbb{R}^{Q_{i} \times 1}}$ and  ${{\bf{W}}^{cc}_i}\in {\mathbb{R}^{Q_{i} \times P_{i}}}$ represent the trainable bias and weight of class $cc$ in layer $i$, respectively. Rectified linear unit (ReLU) is used as the activation function. The input and output feature dimensions of class $cc$ are $P_i$ and $Q_i$, respectively. Similar definitions apply to other classes. 

For class $cc$ in layer $i$, after local feature extraction of user $k$ and reflective unit $n$, the corresponding global output features are computed using \textit{fully connected layer}, while the other classes are processed by applying a weighted average of the results to enhance the network's nonlinear representation. It can be inferred that ${{\bf{f}}_{{kn},i+1}}\in {\mathbb{R}^{4Q_{i} \times 1}}$ and $P_{i + 1} = 4Q_i$ since the output features contain four classes. The whole output feature tensor $F_{i + 1}\in {\mathbb{R}^{4Q_{i} \times K\times N}} $ is three-dimensional, where the feature of the $2{nd}$ and $3{rd}$ dimensional metrics for $k$ and $n$ is ${\bf{f}}_{{kn}, i + 1}$. Thus, good scalability can be anticipated.

For the expansion layer, we employ nine information processing units per class for feature processing.
In the expansion layer, we define the reflective units with CSI as \emph{anchor elements}, as shown in Fig.~\ref{Fig_flowchart}. At first, only $N'=16$ elements constitute the partial CSI.  Each anchor element in an expansion layer is expanded into nine ones by applying the same information processing unit to neighboring reflective units that share the same relative position to the anchor element. Thus, the output includes the features of the original anchor elements, along with those of the adjacent eight reflective units. The output feature of user $k$ and reflective unit $n$, using information processing unit $j$ in layer $i$, is calculated as

	\vspace{-2mm}
\begin{small}
\begin{align}
	\label{expansion layer's feature}
        \begin{gathered}
   {\bf{f}}_{k\nu_{n}^j,i+1}= \hfill\\\left(\begin{array}{c}\text{ReLU}(\mathbf{W}_{i,j}^{cc} \mathbf{f}_{kn,i} + \mathbf{b}_{i,j}^{cc}) \\
\frac{1}{N} \sum_{n'} \text{ReLU}(\mathbf{W}_{i,j}^{ca} \mathbf{f}_{kn',i} + \mathbf{b}_{i,j}^{ca}) \\
\frac{1}{K-1} \sum_{k' \neq k} \text{ReLU}(\mathbf{W}_{i,j}^{oc} \mathbf{f}_{k'n,i} + \mathbf{b}_{i,j}^{oc}) \\
\frac{1}{N(K-1)} \sum_{k' \neq k} \sum_{n'} \text{ReLU}(\mathbf{W}_{i,j}^{oa} \mathbf{f}_{k'n',i} + \mathbf{b}_{i,j}^{oa})\end{array}\right),
    \end{gathered}
\end{align}\end{small}where $\nu_{n}^j$ is the reflective unit index when applying information processing unit $j$ to the input of reflective unit $n$. The index $\nu_{n}^j$ starts from 1 at the top-left corner and increases row by row (i.e., the index for rows $w$ and the columns $h$ is calculated as $h+( w- 1 )H$, where $H$ is the number of columns in the RIS array), then we have
	\vspace{-1mm}
\begin{small}
\begin{align}
  \label{v(n,j)}
\nu_{n}^j = 
\begin{cases} 
n - H - 2 + j & j = 1, 2, 3, \\
n - 5 + j & j = 4, 5, 6, \\
n + H - 8 + j & j = 7, 8, 9.
\end{cases}
\end{align}\end{small}

	\vspace{-2mm}

After each expansion layer, the number of anchor elements is increased to nine times the original amount. In other words, if we have only 16 ($4\times 4$)\footnote{{It is shown that more anchors have higher complexity but are more robust to random channel gain \cite{Peng2025TWC}.}} reflective units with CSI at first,  after two expansion layers as described, the phase shift matrices of 1296 ($36\times 36$) reflective units can be obtained. In addition, to accelerate the training and test processes, we leverage parallel computing by converting the information processing in the dense and expansion layers into tensor operations and using a permutation matrix to sort the RIS reflective units according to \eqref{v(n,j)}.
\end{spacing}
	
	\vspace{-2mm}
	\section{Simulation Results}
	\vspace{-1mm}
\begin{spacing}{0.95}
In this section, we evaluate the proposed algorithm through simulations conducted in Python and PyTorch on a server equipped with an Nvidia GeForce GTX 2080 Ti GPU (11 GB of memory). Besides, we use the open-source DeepMIMO dataset\cite{Alkhateeb2019} to construct the ray-tracing (RT) channels. The {scenario O1\footnote{{An outdoor ray-tracing scenario in the DeepMIMO dataset with two orthogonal streets and one intersection. It includes 18 predefined BS deployed along both streets and a dense grid of user locations indexed by row numbers.}} is chosen, where \texttt{BS 14} and \texttt{BS 6} are selected as fixed BS and RIS, respectively, and user locations are uniformly drawn from rows \texttt{R1000-R1200}.  In this setup, the BS equipped with $9$ antennas and the RIS comprising 1296 reflective units jointly serve $K=4$ users. The training and test sets contain 20480 and 1024 samples, respectively. The transmit power is set to $P_t=1 \rm{W}$. }
During training, we use the Adam optimizer with a learning rate of $8\times10^{-5}$ to update the network parameters, set the batch size to 256, consider ${a_k}=0.25$ for $k \in {\mathcal K}$,  and employ the ReduceLROnPlateau scheduler to adjust the learning rate.

\begin{figure}[t]
	\centering
	\begin{subfigure}
		\centering
        \scalebox{0.31}
		{\includegraphics[width=7.15in]{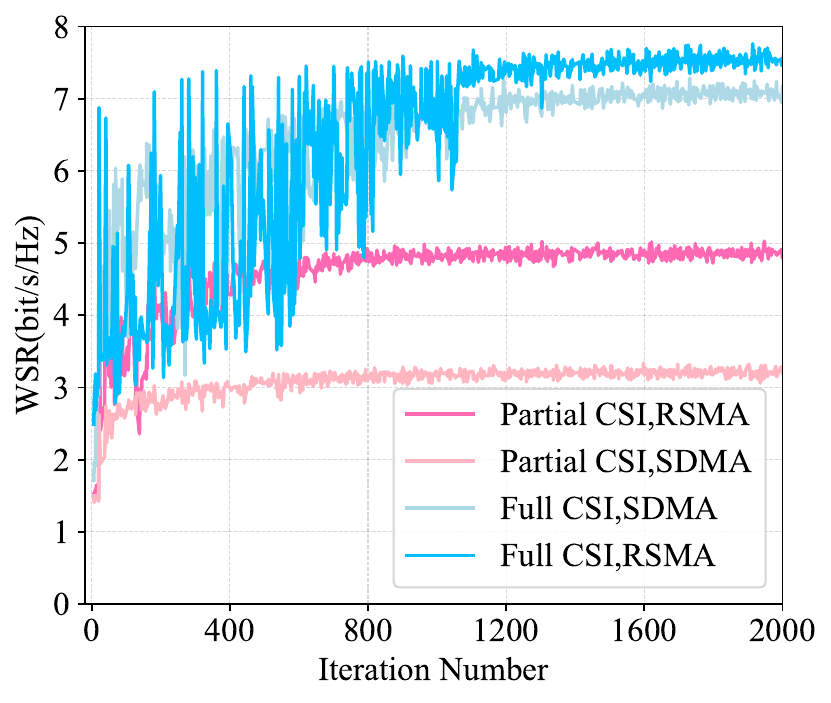}} 
         \vspace{-1mm}
		\caption{The training results of RISnet with SDMA and RSMA under random channels with partial CSI or full CSI as input.}
		\label{Fig_curveiid}
		\vspace{3mm}
	\end{subfigure}
	\begin{subfigure}
		\centering
        \scalebox{0.31}
		{\includegraphics[width=7.15in]{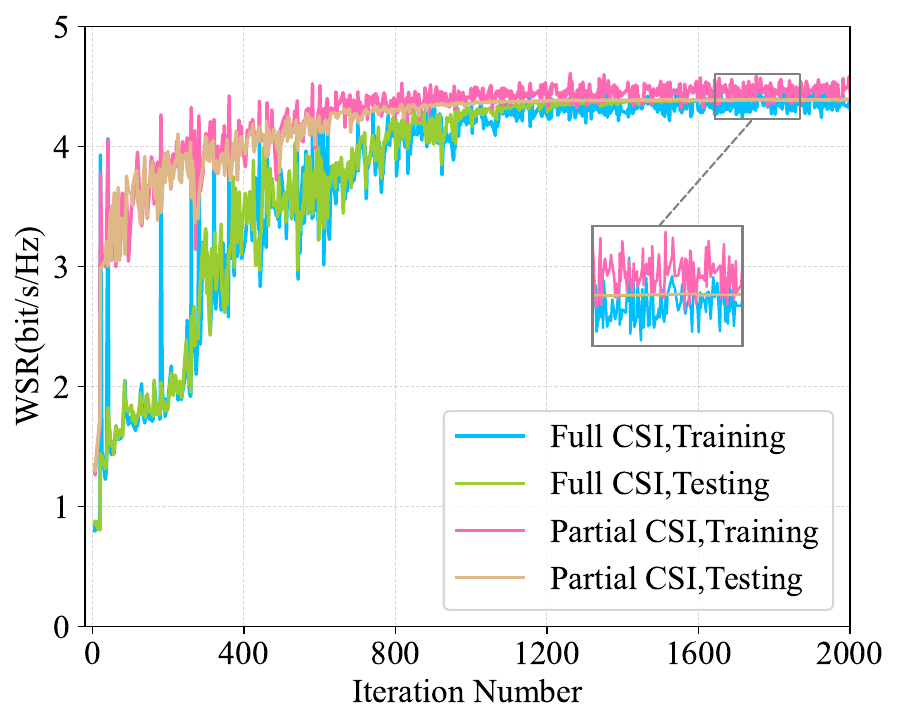}} 
        \vspace{-1mm}
		\caption{The training and testing results of RISnet with RSMA under RT channels with partial CSI or full CSI as input.}
		\label{Fig_curveRT}
		\vspace{3mm}
	\end{subfigure}
	\label{Fig_curve}
    \vspace{-8mm}
\end{figure}

Firstly, in Fig.~\ref{Fig_curveiid}, we compare the performance of RISnet with RSMA and the one using SDMA in the case of rich scattering. The channels to users contain infinite weak multiple path components, which are assumed to be independent and identically distributed complex Gaussian random variables. In this case, the uncertainty is highest when training RISnet with partial CSI. From Fig.~\ref{Fig_curveiid}, we observe that all schemes using partial CSI are worse than those using full CSI, because partial CSI is insufficient to infer the full CSI if the channel is fully random. However, compared with the SDMA scheme, the RSMA counterpart achieves better performance in both full and partial CSI settings. Moreover, the degradation caused by partial CSI over RSMA is much less than that over SDMA, indicating that RSMA improves robustness to partial CSI and mitigates performance loss. It is worth noting that the effectiveness of using partial CSI for RIS configuration depends on the channel model, which can be inferred in Fig.~\ref{Fig_curveRT}. 

\begin{table}[t!]
	\small
	\centering
	\caption{WSR and training time versus the number of layers.} 
	\label{tab:tableI} 
	\begin{tabular}{>{\centering\arraybackslash}p{4cm} 
			>{\centering\arraybackslash}p{0.6cm} 
			>{\centering\arraybackslash}p{0.6cm} 
			>{\centering\arraybackslash}p{0.6cm} 
			>{\centering\arraybackslash}p{0.6cm}} 
		\toprule 
		\textbf{Number of layers }($L$)& 6& 7&\textcolor{purple}{8}& 9\\
		\midrule 
		\textbf{WSR performance} (bit/s/Hz) & 4.283&4.291 & \textcolor{purple}{4.495}&4.221\\
		\midrule
		\textbf{Time of training} (min)& 130.8&131.3&\textcolor{purple}{120.3}&133.2\\
		\bottomrule 
	\end{tabular}
	\vspace{-2.5mm}
\end{table}	

Table I shows the WSR performance and training time versus the number of RISnet layers. It can be observed that, as $L$ increases, the WSR first improves due to enhanced learning capability, but degrades at large $L$ (e.g., $L=9$) because of numerical error accumulation from repeated matrix inversions and multiplications. Thus, $L=8$ provides a good performance–complexity tradeoff. Notice that though training is time-consuming, it is executed only once, and trained RISnet enables fast deployment with low-latency forward inference, ensuring the proposed approach is practical.

Fig.~\ref{Fig_curveRT} presents the performance when applying RISnet with RSMA in training and testing with full and partial CSI under RT channels. It is noted that the RT channel is deterministic, in which case the full CSI can be easily inferred from the partial CSI (since all RIS units share the same propagation paths {with different path loss and complex phase}). It can be observed that training and testing with the same setup achieve similar performance, indicating that no overfitting or underfitting occurs. Similar performance is observed for both full and partial CSI on the RT channel, implying that partial CSI is sufficient for this scenario and that the difficulty of channel estimation is substantially reduced.

In addition to the high performance,
the proposed UL approach requires intensive resource to train the RISnet,
but the inference of a trained RISnet in the application is almost instant,
which allows for real-time deployment with cheap hardware.

\end{spacing}
	
	\vspace{-2mm}
	\section{Conclusions}
	\vspace{-1mm}
	In this paper, we have proposed a RIS-assisted multi-user RSMA downlink system. The WSR maximization problem has been formulated to optimize the precoding matrix and the phase shift configuration. We have solved this problem by involving RISnet training and a low-complexity RSMA precoder. Simulation results have shown that the performance of the proposed scheme approximates that of a full CSI counterpart under deterministic conditions. When channel uncertainty increases, RSMA{-assisted RISnet} has been shown to enhance the robustness, mitigating performance loss significantly.

	%
	%

	\ifCLASSOPTIONcaptionsoff
	
	\fi
    	\vspace{-2mm}
	\bibliographystyle{IEEEtran}
     \begin{spacing}{0.95}
	\bibliography{reference.bib}
    \end{spacing}

	%
	%
	%
	%
	%
	
	
	

\end{document}